# Synthesis of strontium ferrite/iron oxide exchange coupled nano-powders with improved energy product for rare earth free permanent magnet applications

A. D. Volodchenkov [1,2], Y. Kodera [1,2] and J. E. Garay [1,2,3]

[1] Advanced Materials Processing and Sythesis (AMPS) Laboratory
[2] Materials Science and Engineering Program, Mechanical Engineering Department
University of California, Riverside
[3] Materials Science and Engineering Program, Mechanical and Aeronautical Engineering Department
University of California, San Diego

**Abstract**

We present a simple, scalable synthesis route for producing exchange coupled soft/hard magnetic composite powder that *outperforms* pure soft and hard phase constituents. Importantly, the composites is iron oxide based ($SrFe_{12}O_{19}$ and $Fe_3O_4$) and contain no rare earth or precious metal. The two step synthesis process consists of first precipitating, an Iron oxide/hydroxide precursor *directly* on top of $SrFe_{12}O_{19}$ nano-flakes, ensuring a very fine degree of mixing between the hard and the soft magnetic phases. We then use a second step that serves to reduce the precursor to create the proper soft magnetic phase and create the intimate interface necessary for exchange coupling. We establish a clear processing window; at temperatures below this window the desired soft phase is not produced, while higher temperatures result in deleterious reaction at the soft/hard phase interfaces, causing an improper ratio of soft to hard phases. Improvements of $M_r$, $M_s$, and $(BH)_{max}$ are 42%, 29% and 37% respectively in the $SrFe_{12}O_{19}/Fe_3O_4$ composite compared to pure hard phase($SrFe_{12}O_{19}$). We provide evidence of coupling (exchange spring behavior) with hysteresis curves, first order reversal curve (FORC) analysis and recoil measurements.

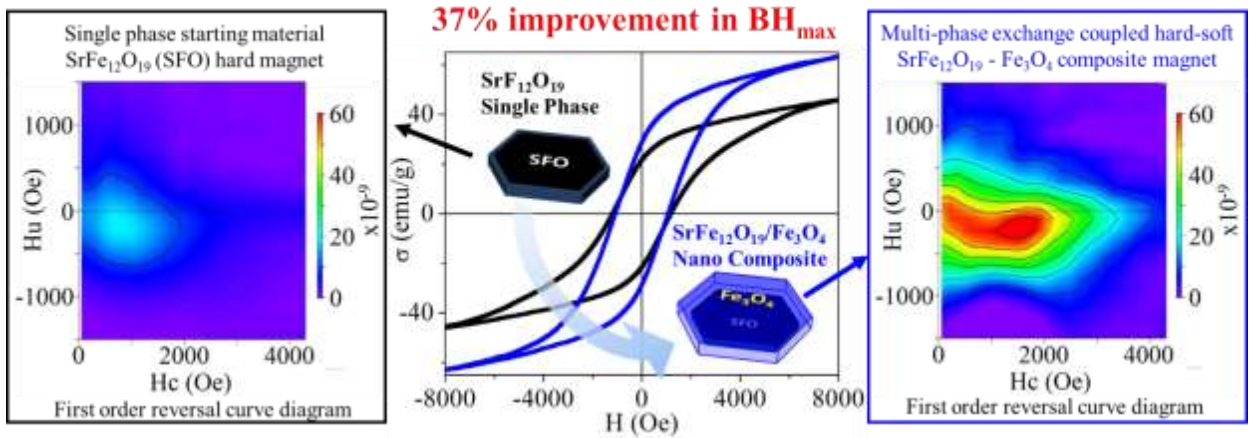

**Introduction and Background**

Permanent magnets (PMs) are essential to an amazing variety of current and future devices, causing widespread interest in improving PM performance. Magnets with high rare earth (RE) content such as Nd-Fe-B (typically with composition $Nd_2Fe_{14}B$), and Sm-Co (typically $SmCo_5$) are currently the state of the art PMs. One approach to increasing PM performance is the exchange spring concept which has been predicted to yield huge gains in energy product, $(BH)_{max}$ (figure of merit for PM performance)[1]. This promise has inspired successful pioneering research in RE based exchanged-coupled PMs[2-6] as well as in Fe-Pt based PMs[7, 8]. For example Lyubinya *et al* [6] showed evidence of exchange coupling in Fe-Pt powders while Liu and Davies[8] showed exchange in RE-iron melt spun alloys.

A natural extension of the exchange spring concept is to replace expensive and threatened magnetic phases (such as RE or Pt based materials) with much more abundant and accessible materials. Along these lines, Debangsu *et al* [9] were able to synthesize all ferrite materials with exchange spring behavior. In this work we chose strontium ferrite ($SrFe_{12}O_{19}$) as a hard phase and $Fe_3O_4$ as a soft phase. $SrFe_{12}O_{19}$, henceforth to be referred to as SFO, is mostly Fe and O which are two of the most abundant elements on earth. This makes SFO a popular PM material in a host of consumer devices.[1,10] While the coercivity of SFO cannot compete with that of RE based hard magnets, the complete elimination of REs and expensive elements is an enticing compromise. The soft phase is composed of only Fe and O and thus is extremely inexpensive and abundant. Cubic $Fe_3O_4$ has a saturation magnetization of 84 emu/g [2], higher than pure SFO (42 emu/g) which can potentially lead to exchange coupled SFO/$Fe_3O_4$ composite that out performs pure SFO.

In addition to using earth abundant, low cost elements, PM material implementation would benefit tremendously from low cost, scalable synthesis procedures. A particular beneficiary would be PMs for motors and generators since kilogram quantities are necessary for these applications.[7] Exchange coupling relies heavily on a high interaction area between the hard phase and the soft phase, requiring the grains of soft and hard material to be in the sub-micrometer/nanometer range and well intermixed[3]. The combination of nano-grains, good mixing, as well as clean interfaces make synthesizing exchange coupled PM using a scalable and economical synthesis procedure challenging.

Among the many different approaches to synthesize powder, the homogeneous precipitation (HP) method is one of the most common approaches used in inorganic chemistry for laboratory and industrial scales[11, 12]. HP generally provides good morphology and crystal phase control without the need for extreme conditions/systems, such as rapid heating/cooling, high vacuum, high pressure, etc. There have been previous successful soft chemistry routes[13] for obtaining exchange coupled particles, although these pioneering cases did not lead to improved $(BH)_{max}$. Our concept for synthesizing composite PMs is a twostep process, shown schematically in **Figure 1**. First we use HP to precipitate a Fe-O/Fe-O-H precursor *directly* on top of SFO nano-flakes, ensuring a very fine degree of mixing between the hard and the soft magnetic phases. We then use a second step that serves to reduce the precursor to create the proper soft magnetic phase and create the intimate interface necessary for coupling. The result is a simple, inexpensive synthesis route for exchange coupled PM composite powder that *outperforms* pure SFO.

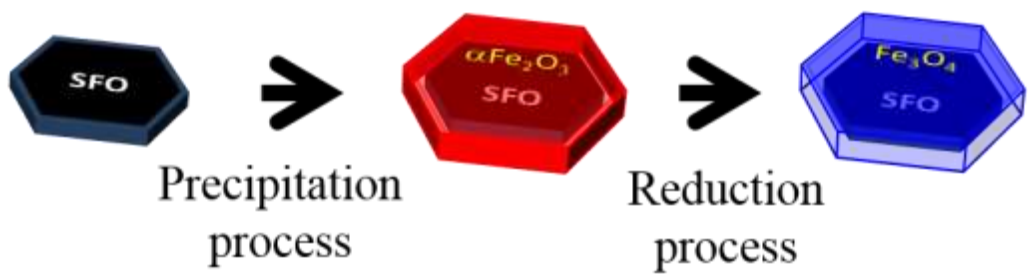

*Figure 1. Schematic of hard -soft composite synthesis.*

**Experimental procedure**

*a. Material synthesis*

Precipitation by decomposition of urea was chosen as the method of depositing Fe oxide/hydroxide (Fe-O/Fe-O-H as the precursor of the soft phase) onto SFO powder ($SrFe_{12}O_{19}$, Nanostructured & Amorphous Materials Inc). The SFO powder consists of high aspect ratio flake like particles with average diameter of 1.12 μm and thickness of 0.16 μm. As starting materials for Fe oxide/hydroxide precipitation, 28.7 mmol of $Fe(NO_3)_3$ (Sigma Aldrich >98%) and 167 mmol of $CO(NH_2)_2$(Urea; Sigma Aldrich >99.5% )were mixed into 150 ml of $H_2O$. The mixture was then titrated into a slurry of 1.58 mmol of SFO and 50 ml of $H_2O$. The temperature was maintained at 90 ºC for 2 hours. The resulting composite powder was cooled quickly to prevent further particle growth.

In order to understand the Fe-O/Fe-O-H precipitation process (phase and morphology development during the process) we also did experiments without the SFO (soft phase only) in which no titration is used. The same amount of Fe(NO$_3$)$_3$ and 167 mmol of Urea were mixed into 150 ml of H$_2$O and held at 90 ºC for 2 hours. The resulting homogeneous precipitation creates nano-scale particles of the soft magnetic phase precursor.

In both the composite and soft phase cases, the resulting particles were centrifuged, washed with ultra-high pure water and centrifuged again. The powder and liquid was separated by decantation. The powder was dried at 80 ºC in a vacuum furnace for 24 hours to ensure no moisture remains after which the dried agglomerates are broken by mortar and pestle. The powders were then treated in a tube furnace under forming gas (5% H$_2$, 95 % N$_2$) at temperatures ranging from 300ºC to 500ºC with 1 hour ramp and no hold time. The resulting powder was handled in Argon atmosphere to avoid oxidation.

The yield of Fe-O precipitation was obtained by the gravimetric analysis based method. The collected liquid from decantation was dried and calcined at 800 ºC for 6 h in air atmosphere. The residue remaining after calcination was α-Fe$_2$O$_3$ single phase which was confirmed by X-ray diffraction analysis. The amount of α-Fe$_2$O$_3$ was used to calculate the yield of the precipitation process.

*b. Structural and microstructural characterization*

The phase composition was characterized with X-ray diffraction (XRD) (PANalytical Empyrean Diffractometer with Cu Kα X-ray source $\lambda_{K\alpha 1}$=1.54056 Å $\lambda_{K\alpha 2}$=1.54440 Å using 0.01313º step size). In order to provide a simplified estimate of phase composition ratio, XRD peak intensities ratio was calculated by taking the highest intensity peak of one phase and dividing it by the sum of the highest intensity peaks of all detectable phases and multiplying by 100. The particle morphology was characterized by Scanning Electron Microscopy (SEM) (Philips XL30).

*c. Magnetic measurements*

Magnetic properties were measured using a Vibrating Sample Magnetometer (VSM) (Lakeshore 7400 Series) at room temperature. Hysteresis loop measurements using field values of up to 1.7 T were obtained in order mass normalized magnetization, σ [emu/g] *vs.* applied field, H [Oe]. We refer to these measurements as customary hysteresis loops. Coercivity, $H_c$ [Oe], remanence magnetization $M_r$ [emu/g] and saturation magnetization $M_s$ [emu/g] was extracted from the σ *vs.* H hysteresis curves. In calculating $M_s$, the non-saturating slope (due to SFO and $Fe_3O_4$ being ferrimagnetic) was subtracted. Energy product, $(BH)_{max}$ [MGOe] was calculated assuming full density of SFO, .

First order reversal curve (FORC) measurements were done by ramping the field to 0.6 T then decreasing the magnetic field to a reversal field with value of $H_a$ and ramping back up to 0.6 T through field values, $H_b$ with a step size of 200 Oe. Magnetization as a function of $H_a$ and $H_b$, M$(H_a,H_b)$, is recorded. This procedure was repeated in order to measure a collection of first order reversal curves for reversal fields in 200 Oe intervals from 5800 Oe to -6000 Oe. FORC distribution, $\rho$ is calculated using the relation[5]:

$$\rho(H_a, H_b) = -\frac{\partial^2 M(H_a, H_b)}{\partial H_a \partial H_b} \qquad (1)$$

FORCinel was used to calculate the FORC distribution and plot it, traditionally, as $H_c$ *vs.* $H_u$ where $H_c = (H_b-H_a)/2$ and $H_u = (H_a+H_b)/2$ [14].

Recoil loop measurements were done by first ramping magnetic field to 1.7 T in order to bring the sample to saturation (ignoring the non-saturating component due to SFO and $Fe_3O_4$ being ferrimagnetic). A Reversal field, $H_a$, was applied, removed (field taken to 0 Oe) and reapplied (field taken back to $H_a$). $H_a$ values were varied from 100 Oe to 1400 Oe in 100 Oe increments. Measurements of magnetization, σ, were taken from $H_a$ to 0 Oe, forming the recoil magnetization curve, and likewise from 0 Oe back to $H_a$, forming the recoil demagnetization curve. The area between recoil magnetization and recoil demagnetization curves was calculated using numerical methods. Normalized recoil loop area was calculated by dividing the area between recoil magnetization and recoil demagnetization curves by one half of the total area of the sample's customary hysteresis loop area (also calculated using numerical methods). $M_{recoil}$ is the value of magnetization with 0 Oe field applied following recoil magnetization from $H_a$ to 0 Oe. Recoil remanence ratio $M_{recoil}/M_r$ was calculated by taking the $M_{recoil}$, values and dividing it by the magnetic remanence $M_r$ (from the customary loop obtained as described above).

**Results and Discussion**

*Synthesis of soft magnet phase*

We start by discussing the Fe-based soft magnet powders (Fe-O/Fe-O-H) *i.e.* materials without the hard SFO phase. Although many different approaches have been reported for synthesizing iron hydroxide/oxides[11, 12, 15], it is important to understand the detail of synthesis and reduction behavior of iron hydroxide/oxide in our experimental conditions in order to achieve our goal of obtaining soft-hard composites with controlled properties. The precursor was synthesized through precipitation by thermal decomposition of urea in iron nitrate solution[12,16]. There are several reports on the synthesis of iron hydroxide/oxide and oxide using iron salt and urea as reactant that produce fine nano-particles and high yields[16-19]. We chose iron nitrate and urea as reactants in this particular study because of the simplicity of the removal of ammonium nitrate (formed as byproduct) in the process.

**Figure 2** shows XRD patterns of the as-precipitated powder from iron nitrate and urea solution heated at 90 °C. The XRD confirms that the product is mixture of α-$Fe_2O_3$ and α-FeO(OH) and low crystallinity which agrees with previous results[20]. The high background intensity at low angles suggests the presence of low crystallinity/amorphous phase.

In order to achieve the desired soft phase ($Fe_3O_4$) the dried as-precipitated powder was then thermally treated in 95% $N_2$: 5% $H_2$ gas flow. The XRD in **Fig 2** shows the influence of treating temperature on the reduction of the precipitated α-$Fe_2O_3$/α-FeO(OH). Increasing temperature causes decomposition of α-FeO(OH) and reduction of hexagonal $Fe_2O_3$ to cubic $Fe_3O_4$ and subsequently reduction of $Fe_3O_4$ to metal α-Fe at higher temperature. These observations agree with previous studies[21-24]. FeO was not detected during the reduction process because the processing temperature was kept below 570 ℃[25]. The phase evolution is more easily appreciated in **Figure 3** showing XRD peak intensity ratio vs. reduction temperature. At the processing temperature of 300 °C, all of the as-precipitated phases are converted to $Fe_3O_4$. The metal α-Fe appears at 350 °C and its relative amount increases with temperature, however it is not the major component until reduction temperature is increased to 450 ºC. The conversion ratio reaches over 80% at 500 ºC based on simplified estimation from XRD intensity ratio.

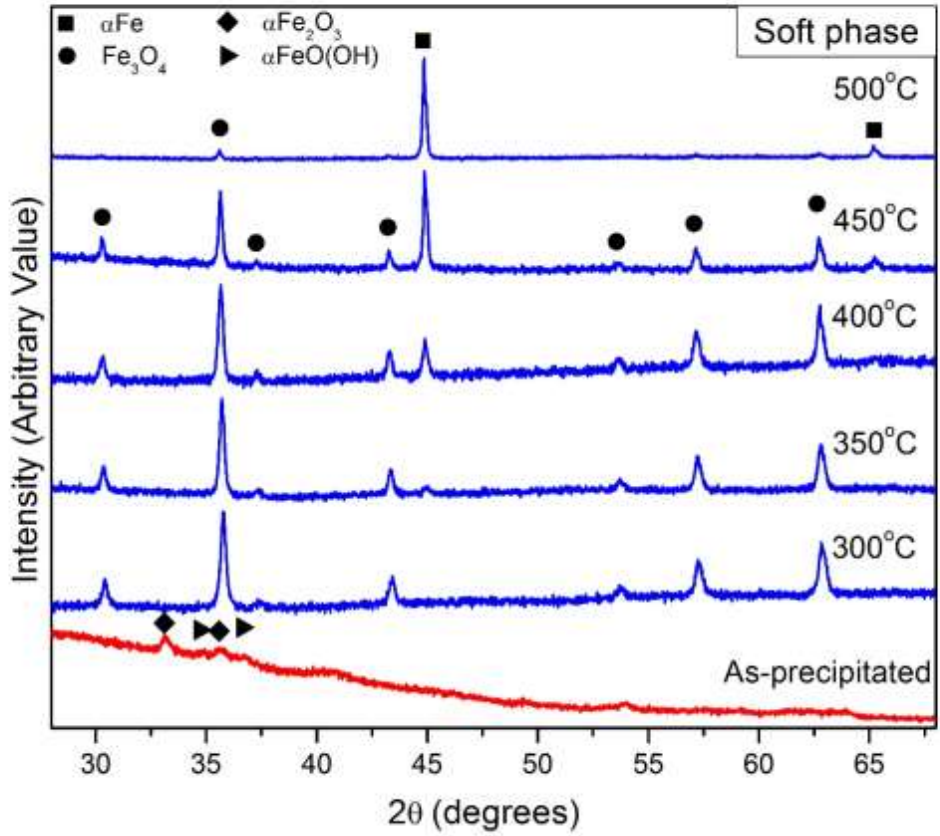

*Figure 2: X-ray diffraction patterns for the soft phase after precipitation as well as at reduction temperatures 300-500ºC.*

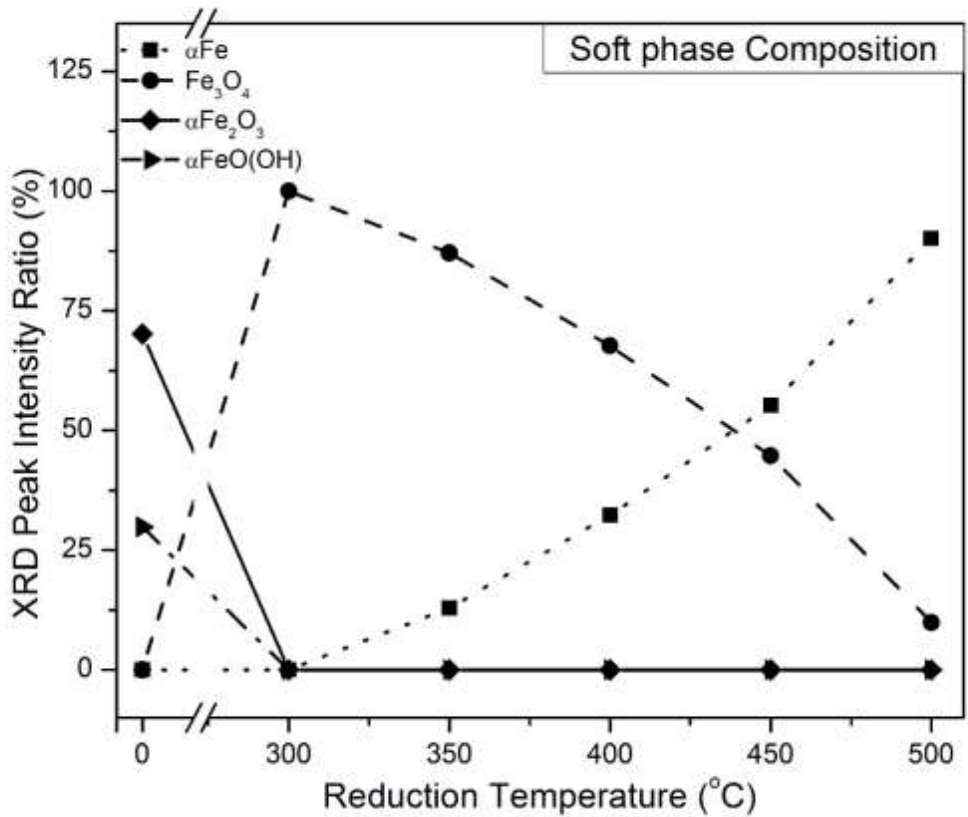

*Figure 3: XRD peak intensity ratios at various reduction temperatures. The peak intensity ratio is the ratio of the most intense peak of a particular phase to the sum of the intensities of the most intense peaks of all identifiable phases. The most intense peaks for α-Fe, $Fe_3O_4$, α-$Fe_2O_3$ and α-FeO(OH) are from the (110), (311), (104) and (101) planes, respectively.*

**Figure 4** shows SEM micrographs of as-precipitated powder as well as heat treated ones. The as-precipitated particles exhibit morphology with low aspect ratio and small grain size (tens of nm). Also the micrographs suggested that the low degree of aggregation of particles. These morphological characteristics contribute to the ease of reduction.

The SEM micrographs of sample reduced at 400 °C shows significant change in surface roughness and evidence of grain growth. Considering this result and the reduction mechanism, we assume about 70% of $Fe_3O_4$ remains as a core that is surrounded by α-Fe (about 30%) as a shell at 400 ºC. Despite the clear grain growth, these soft phases did remain in the nanoscale which is very important for obtaining an exchange coupled PM.

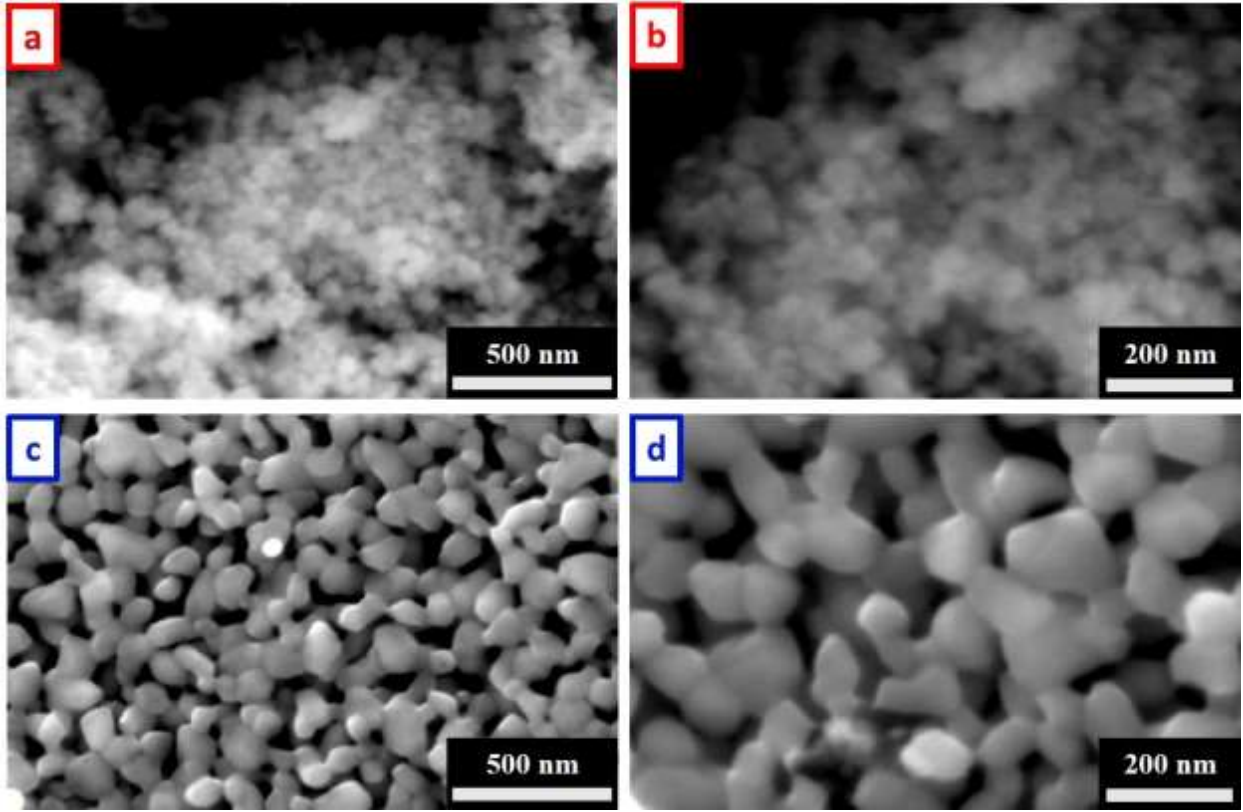

*Figure 4: (a), (b): SEM micrographs of as-precipitated soft phase powder. (c), (d): SEM micrographs soft phase powder reduced at 400ºC.*

*Synthesis of soft/hard magnet composites*

**Figure 5** shows XRD of SFO powders after having undergone the precipitation of soft phase precursor. Also shown in **Figure 5** are the XRD patterns of the soft/hard composites powders after reduction at various temperatures. The as-precipitated sample shows XRD patterns identical to as-received SFO (not shown here) and no significant peak from precipitate except small peak at 33 degrees corresponding to α-$Fe_2O_3$. This result suggests that there is no significant damage of SFO during the precipitation process. **Figure 6** shows XRD peak intensity ratio vs. reduction temperature. At 300 °C, the intensity of α-$Fe_2O_3$ peaks increase and the peaks of $Fe_3O_4$ appear. By contrast, the soft phase only results (**Figure 3**) show the full conversion of precursor to $Fe_3O_4$ at same temperature (300 °C). In the case of non-composite system, the entire surface of particle, except point contacts between particles, is exposed to atmosphere, on the other hand, in the composite case (**Figure 6**), α- $Fe_2O_3$ particles are precipitated on the SFO meaning at least one side of particles is not exposed to atmosphere; this should slow down reduction kinetics of $Fe_3O_4$ formation

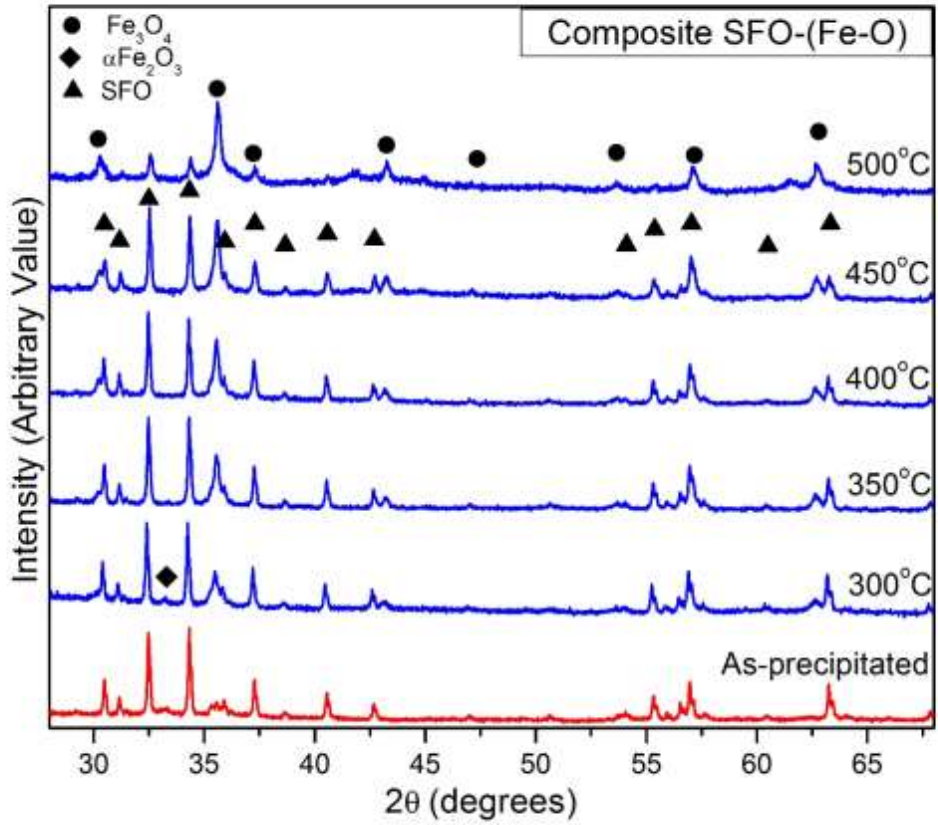

*Figure 5 X-ray diffraction patterns for the SFO-(Fe-O) composite after precipitation procedure as well as at reduction temperatures 300-500°C.*

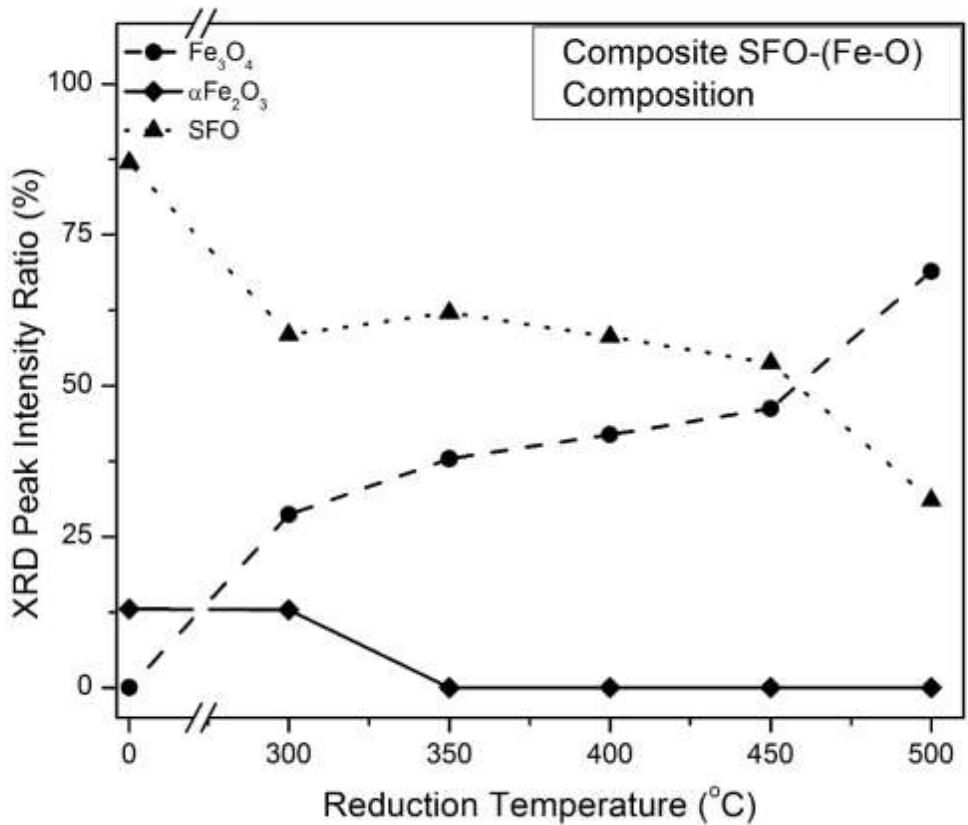

*Figure 6: XRD peak intensities ratios at varying reduction temperatures. The peak intensity ratio is the ratio of the most intense peak of a particular phase to the sum of the intensities of the most intense peaks of all identifiable phases. The most intense peaks for Fe$_3$O$_4$, α-Fe$_2$O$_3$ and SFO are from the (311), (104) and (107) planes, respectively.*

In addition to reaction of the precipitated layer the decay of SFO was simultaneously observed by XRD. At 500 °C, the Fe$_3$O$_4$ peaks become more intense than the SFO peaks. This suggests that the reaction between the deposited layer and SFO is taking place. This reaction is more intense at higher temperature. Based on the results of the soft phase study in (**Figure 3**), the heat required to reduce iron oxide to metal α-Fe is enough to cause significant reaction between SFO and soft magnet phase. In other words, too low a temperature produces SFO/α-Fe$_2$O$_3$ composites instead of the desired SFO/Fe$_3$O$_4$ composites, while too high a temperature destroys the SFO phase, resulting in a composite with too much soft phase. This data clearly show that there is a *limited processing temperature window* that produces the desired phases.

**Figure 7** shows SEM micrographs of samples of the as-received SFO powder as well as composite powders after the precipitation step and after the reduction step at various magnifications. The As-received SFO (**Figure 7 a-c**) exhibits hexagonal facets which correspond to its crystal structure. When the precursor is precipitated in SFO suspension, there are "clouds" of particles deposited on the surface and at the intersections of SFO grains (**Figure 7 d-f**). The higher magnification images confirm that those particles size are tens of nanometers.

The calculated yield of the precipitated material was 62%. When precipitated the Fe based soft phase forms only $Fe_3O_4$ (e.g. the sample reduced at 400 °C), the volume ratio of hard /soft phase (SFO /$Fe_3O_4$) is 55/45. Therefore, the estimated thickness of $Fe_3O_4$ coating is calculated to be 37 nm on the surface of SFO assuming a hexagonal plate-like particle as shown in **Figure 1**. The dimensions used for this estimate are the average sizes measured for SFO (diameter of 1.12 μm and thickness of 0.16 μm).

After reducing composites at 400 °C, there is not significant change in low magnification images and there are still relatively sharp corner/edge of SFO composite grains. This suggests that sintering and grain growth of SFO composite is not significant at this temperature. However, high magnification images (compare between **Figure 7 a, d** and **g**)) reveal a clear change in the surface morphology of SFO composite. The SFO composite grains show agglomerated round shaped particles (tens of nanometers) with curved edges and many pores. This is caused by the reduction of precursor and sintering of $Fe_3O_4$ during the reduction process. First, both decomposition and reduction reduce the number of atoms in the compound and create voids and pores. Then, sintering forms neck and causes grain growth. Therefore, those voids and pores are segregated from the matrix to form the microstructure we can see in **Figure 7i.**

There is a clear contrast between the soft phase only and composite powders in the degree of grain growth. When the samples are treated at 400 ºC, the soft magnet shows significant grain growth, on the other hand, the composite (soft phase on SFO) does not show such a drastic growth (see **Figure 4c** and **Figure 7i**). In order to have sufficient significant grain growth, two conditions are required: 1. There needs to be sufficient atomic mobility for grain growth. 2. There needs to be sufficient material to form coarse grains with larger volume. Since the soft phase results (**Figure 4**) show significant grain growth there is sufficient thermal energy at 400 ºC required for mobility indicating that there simply is not enough soft phase on the surface of SFO to form large grain in the composite case.

These structural and morphological characterizations confirm that we successful in synthesizing SFO/Fe$_3$O$_4$ nanocomposites by precipitation of precursor on the SFO followed by a reduction process.

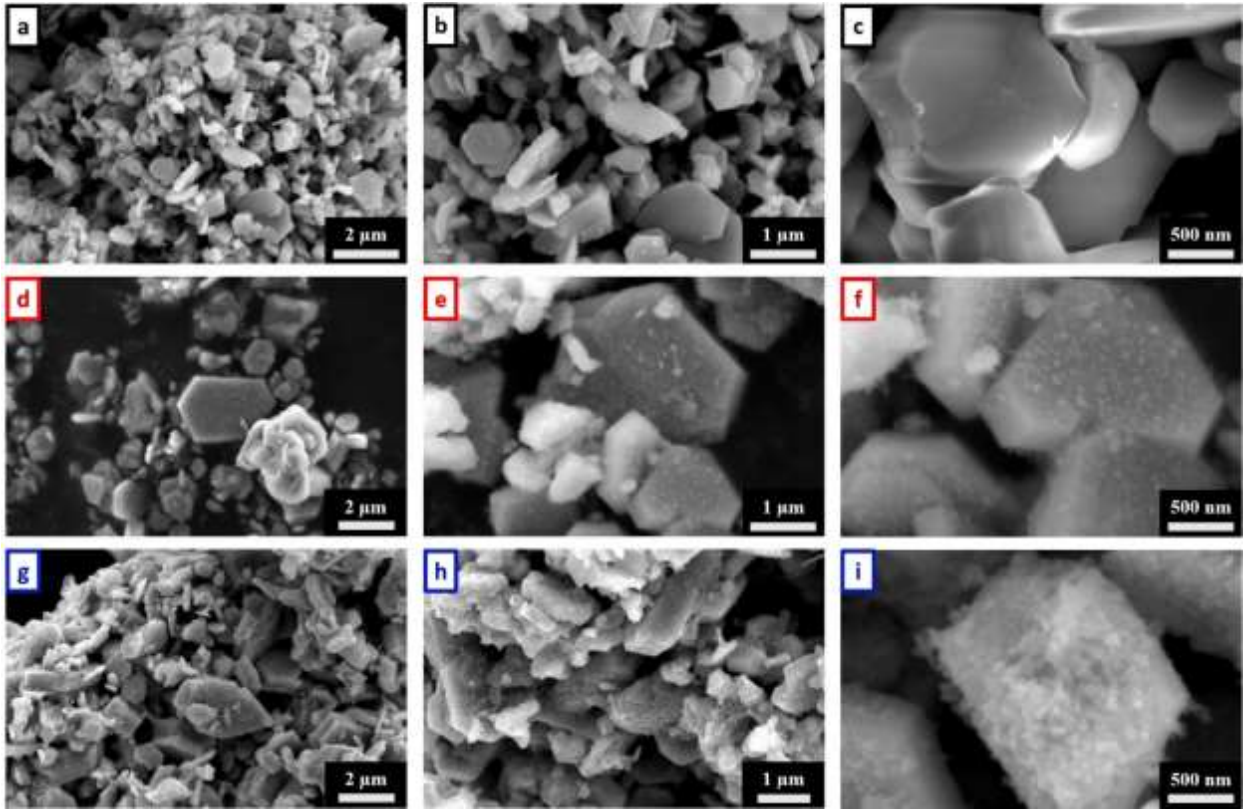

*Figure 7: (a), (b), (c): SEM micrographs of single phase SFO powder. (d), (e), (f): SEM micrographs of composite powder after precipitation procedure. (g), (h), (i): SEM micrographs of composite powder after reduction at 400 ºC.*

*Magnetic Properties*

Magnetic saturation, $M_s$ taken from the measured hysteresis loops of soft phase powders is displayed in **Figure 8.** Post precipitation, (0 ºC in **Figure 8**) the powder has very low $M_s$ as expected for a powder composed of α-Fe$_2$O$_3$ and α-FeO(OH) (see XRD results (**Figure 2**)). Both α-Fe$_2$O$_3$ and α-FeO(OH) are antiferromagnetic. As we reduce the powder at 300 ºC the hexagonal α-Fe$_2$O$_3$ becomes mostly cubic Fe$_3$O$_4$ with saturation magnetization of 93 emu/g. $M_s$ increases further after 400 ºC as the composition of α-Fe increases and cubic Fe$_3$O$_4$ decreases. After reduction at 500 °C the powder has a very high $M_s = 147$ emu/g as expected for a powder that is mostly α-Fe.

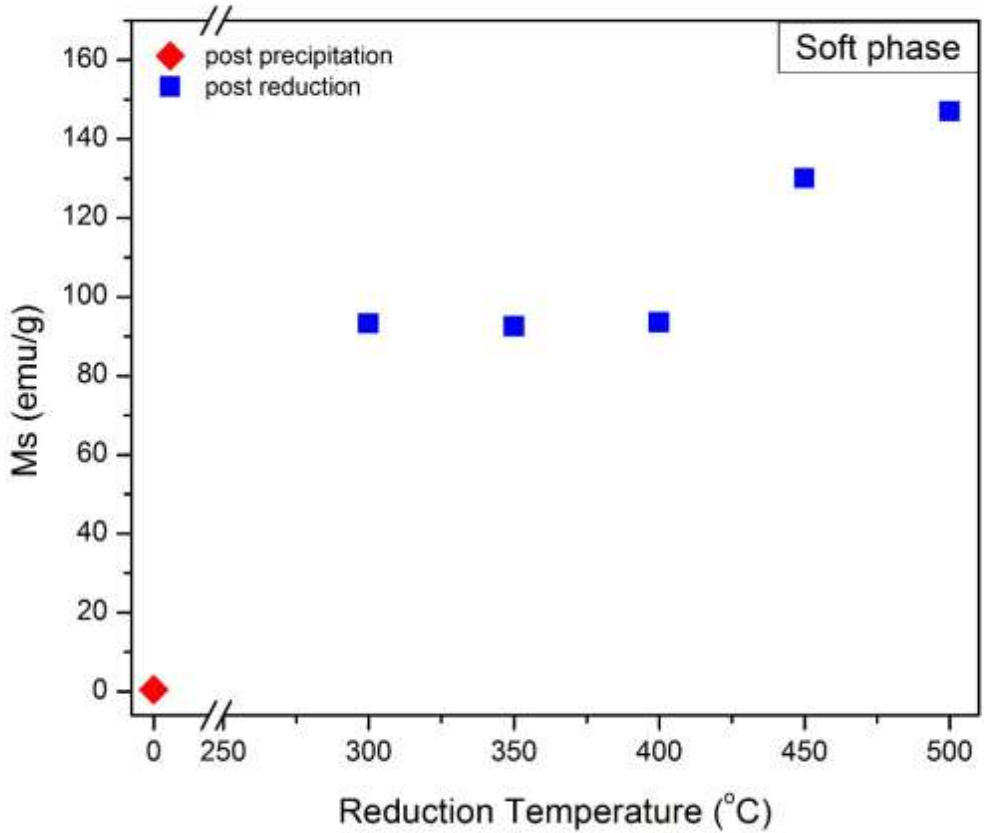

*Figure 8: Saturation magnetization of the soft phase at various reduction temperatures.*

Magnetic properties of the as received SFO and composite powder are displayed in **Figure 9**. The single phase SFO powder has coercivity, $H_c$, remanence magnetization, $M_r$, saturation magnetization, $M_s$ and energy product, (BH)$_{max}$ values of 967 Oe, 18.3 emu/g, 42 emu/g and 0.165 MGOe. The (BH)$_{max}$ values were values calculated assuming full density of 5.1 g/cm$^3$ for SFO[26]. These values provide bench mark values *i.e.* the main goal of the study is to achieve a composite PM with (BH)$_{max}$ higher than 0.165. It should be noted that it is possible to produce SFO magnets with higher (BH)$_{max}$, but optimization of magnetic performance typically requires grain alignment to achieve highest coercivity. Here we are comparing random (unaligned) magnetic powders in both the composites and pure SFO cases.

After initial precipitation, the composite powders show a decrease in magnetic properties This is not surprising because the addition volume of antiferromagnetic α-$Fe_2O_3$ dilutes SFO's remanence, saturation and energy product. The magnetic properties are improved with reduction however; **Figure 9** also shows that magnetic properties have a clear dependence on the reduction temperature.

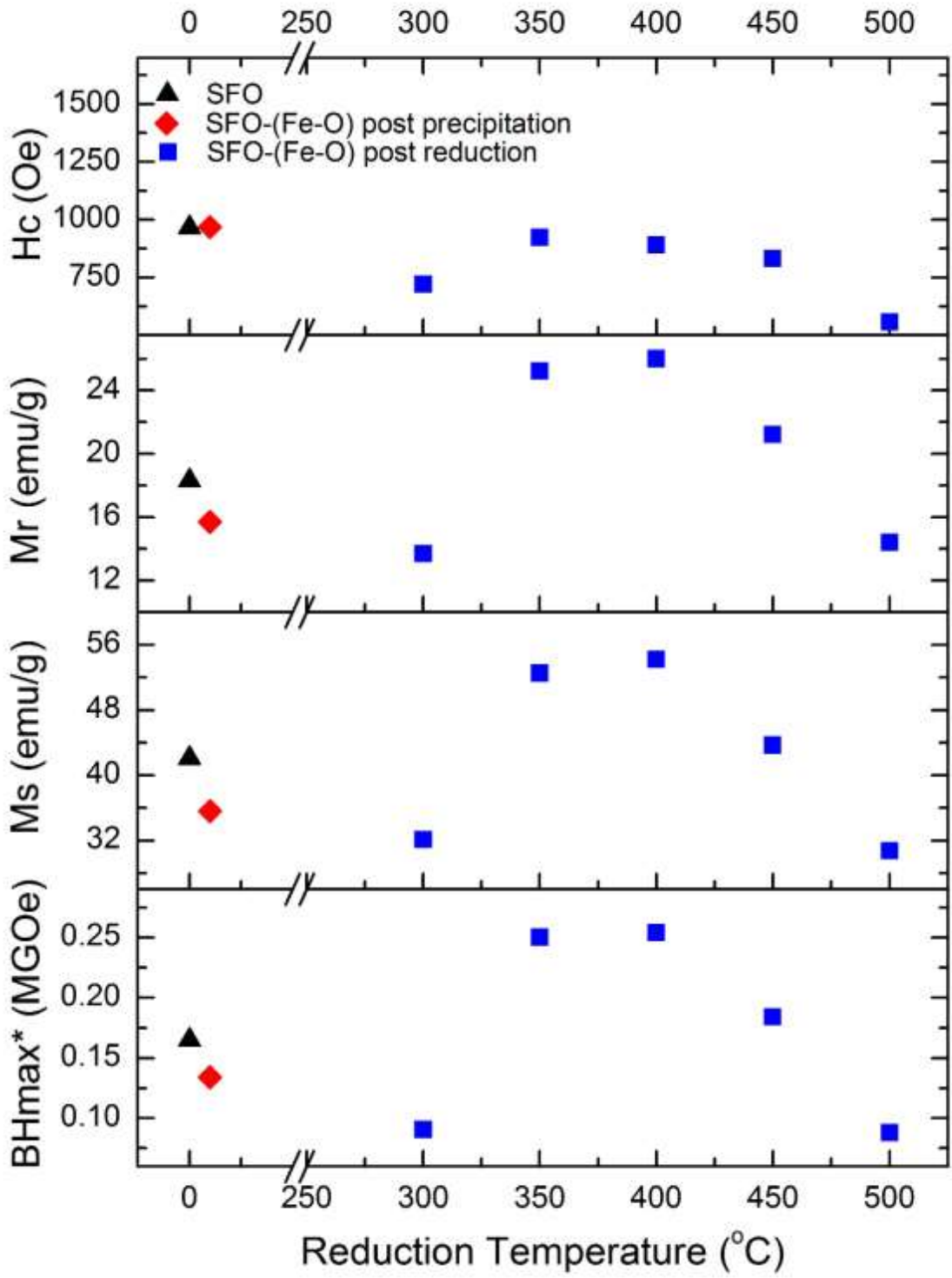

*Figure 9: SFO/(Fe-O) composite magnetic properties (coercivity, remnant magnetization, saturation magnetization, energy product).*

As the reduction temperature increases, the magnetic properties improve significantly, at intermediate temperatures (250 - 400 °C) and decrease again at 450 °C. This "optimal" temperature finding mirrors the processing window effect shown for the phase evolution (Section IIIb) and can be explained as follows. At low temperatures the soft phase has not been fully converted to the desired soft ferrimagnetic $Fe_3O_4$. At reduction temperatures higher than 400 °C there is too much reaction between the SFO and the precipitated phase, causing there to be too much soft ferrimagnetic phase relative to hard ferrimagnetic phase. The maximum $(BH)_{max}$ is achieved by reducing the composite at 400 °C. Notably, all magnetic properties, except coercivity, of this nanocomposite powder surpass those of the pure SFO powder. Improvements in $M_r$, $M_s$, and $(BH)_{max}$ are 42%, 29% and 37% respectively. As noted earlier, it is likely that the $(BH)_{max}$ of these composites can be further increased by aligning the magnetic phases. This should be facilitated by the flake-like nature of our SFO and composites phases. Further studies in this direction are underway.

**Figure 10** compares hysteresis loops of the pure SFO and $SFO/Fe_3O_4$ composite reduced at 400 °C. The smooth hysteresis curve for the composite is indicative of single phase behavior and therefore is evidence that the hard and soft phases are exchange coupled[27]. If the composite was decoupled, the curve would have a clear kink resulting from the hard-soft phases acting independently.

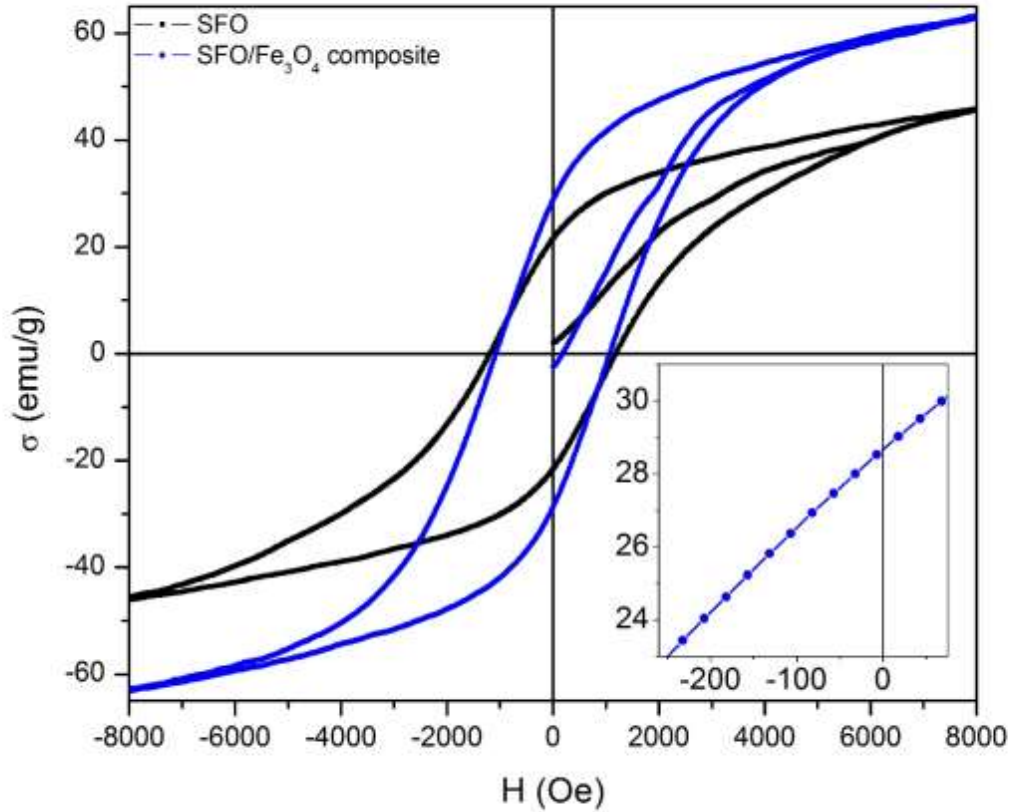

*Figure 10: SFO/Fe$_3$O$_4$ composite (reduced at 400°C) hysteresis loop compared to single phase SFO.*

FORC diagrams are useful for gauging particle interactions of magnetic particles. **Figure 11** shows the FORC diagram for SFO/Fe$_3$O$_4$ composite (**Figure 11b**) as well as pure SFO hard phase (**Figure 11a**). Comparison of the two diagrams shows clear differences. The maximum value for the FORC distribution is $65 \times 10^{-9}$ for the composite material and $19 \times 10^{-9}$ for the SFO. Generally, a higher maximum value indicates more ferromagnetic interactions[28, 29], which we attribute to the existence of the soft phase in the composite.

There is also a larger spread of $H_u$ data, suggesting more particle interactions (larger mean interaction field) in the composite material. The location of the density distribution peak *i.e.* 'density hotspot' can also help interpret the nature of interactions. A hotspot located below the $H_u = 0$ axis further indicates interacting particles[30] and is characteristic of an exchange style interaction[28]. The hotspot peak is shifted to $H_u$ = -236 Oe for the composite which strongly corroborates the hysteresis curve results (**Figure 10**) demonstrating exchange coupling.

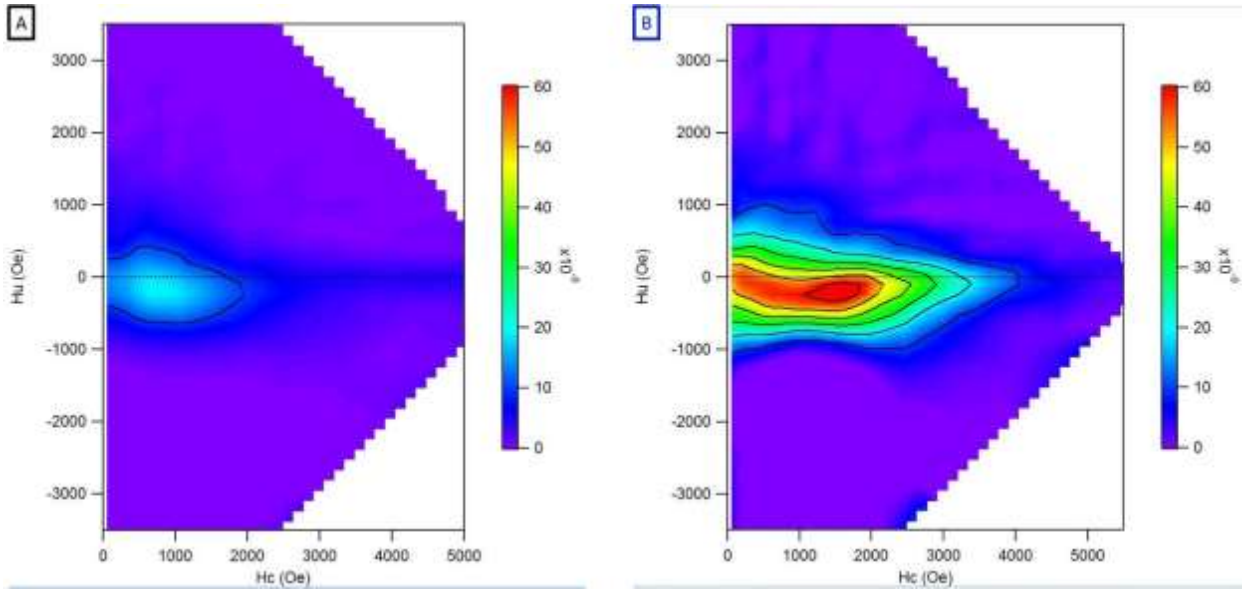

*Figure 11: First order reversal curve (FORC) diagrams for (A) single phase SFO and (B) SFO/Fe$_3$O$_4$ composite (reduced at 400°C).*

To further confirm the existence of coupling in our composites, we did recoil loop measurements; recoil loop analysis has been used previously to evaluate coupling in nanocomposite permanent magnets[31-33]. **Figure 12** shows recoil loop measurements for SFO/Fe$_3$O$_4$ composite as well as pure SFO hard phase, along with the recoil loop areas and recoil remanence for both SFO/Fe$_3$O$_4$ composite and pure SFO.

In a single phase (or very well coupled) magnet, one would expect closed loops *i.e.* little to no area between the magnetization and demagnetization curves. Open recoil loop are often attributed to partial or total decoupling of the soft phase and hard phases in nanocomposite magnets [34] Although not typically expected in a single composition magnet, the pure SFO exhibits open recoil loops (**Figure 12a**). Open recoil loops have been reported before in nano-scale single composition magnets and attributed to inhomogeneity in magnetic anisotropy[35], thermal fluctuation[36] and intergranular exchange interactions[37]. We believe that one or more of these cause open loops in pure SFO. Comparison of the 'openness' of the curves in **Figure 12a** and **12b** reveals a very similar recoil behavior in SFO/Fe$_3$O$_4$ composite and SFO.

A more quantitative comparison of the recoil loop area can be achieved by normalizing the recoil loop are by ½ of the total hysteresis loop area. The data for SFO/Fe$_3$O$_4$ composite in comparison to pure SFO is plotted in **Figure 12c**. The fact that the open area is almost identical means that the addition of the soft phase Fe$_3$O$_4$ is not increasing the degree of decoupling of the composite compared to SFO. This is not surprising in light of the hysteresis curve and FORC analysis discussed above, showing that the SFO/Fe$_3$O$_4$ composites behave as exchanged coupled magnets.

From a permanent magnet development point of view, the primary reason one wants an exchange spring behavior is to increase the energy product (BH)$_{max}$ as we show here. Another very important benefit is to increase the magnet's resistance to demagnetization. A measure of resistance to demagnetization can be obtained by calculating the ratio of the remanence value measured during a recoil measurement, $M_{recoil}$ (see experimental procedure for details) to the $M_r$ measured during a standard hysteresis loop. This ratio is plotted for the nanocomposite and SFO in **Figure 12d**. An exchange coupled nanocomposite magnet with optimal microstructure has a partially reversible demagnetization curve[38], meaning that at low H$_a$ values, the $M_{recoil}/M_r$ ratio is near 1. We see that this is the case for the recoil remanence ratio (**Figure 12d**) of the nanocomposites. The data also reveal that the $M_{recoil}/M_r$ values are higher for the SFO/Fe$_3$O$_4$ composite that the pure SFO at all applied magnetic fields, demonstrating that the composites are more resistant to demagnetization and providing further evidence of exchange coupling.

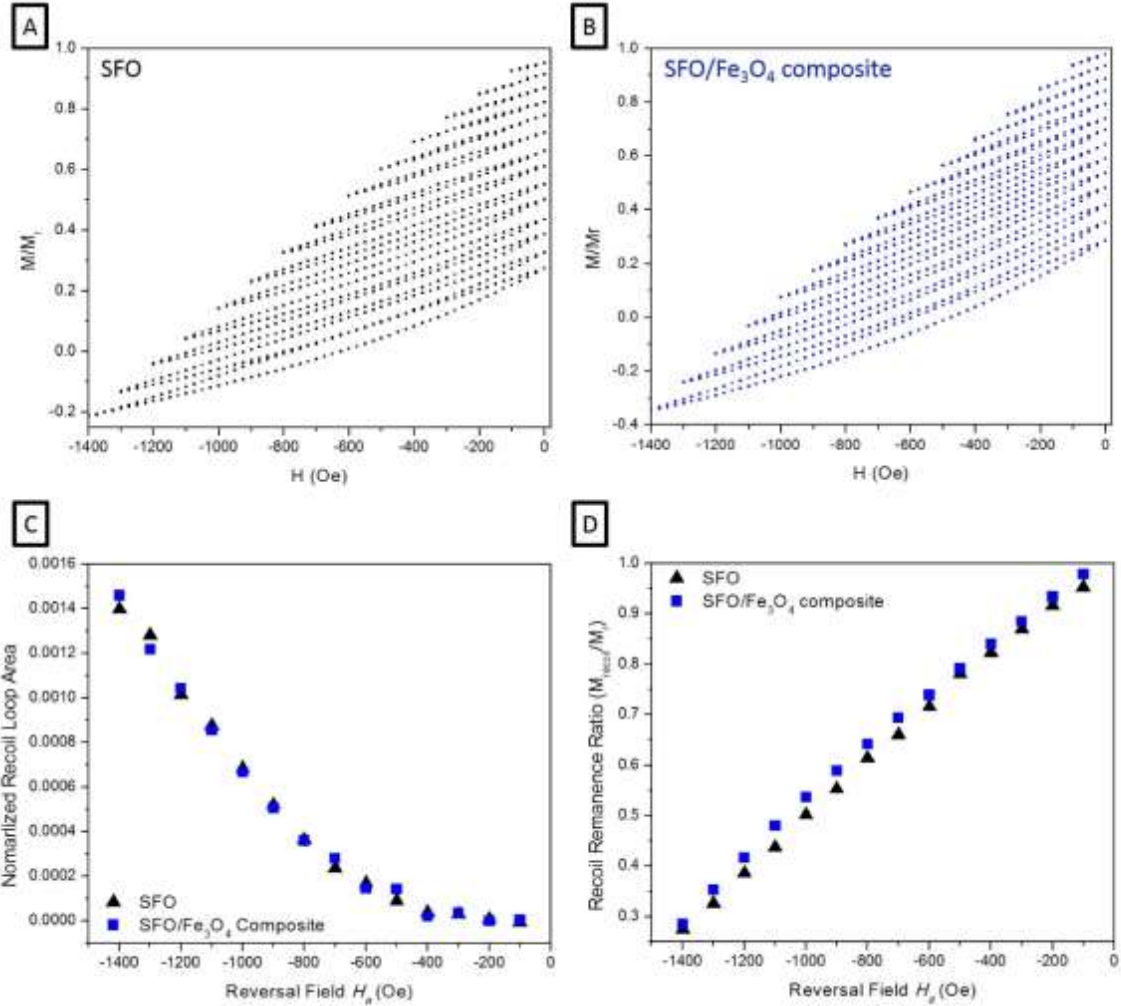

*Figure 12: (A) Single phase SFO recoil loop measurement. (B) SFO/Fe₃O₄ composite (reduced at 400°C) recoil loop measurement. (C) Normalized recoil loop areas (normalized by ½ full hysteresis area) for pure SFO and SFO/Fe₃O₄ composite (reduced at 400 °C). (D) Recoil remanence ratio for pure SFO and SFO/Fe₃O₄ composite (reduced at 400 °C).*

**Summary**

In summary, we have presented a synthesis and processing procedure for the production of SFO/Fe$_3$O$_4$ exchange coupled nanocomposites that contain no rare earth or precious metals. Our procedure is a simple scalable procedure that relies on a precipitation step followed by a reaction step. The precipitation step ensures intimate contact and good intermixing of the two phases. The reaction step allows for the conversion of the as-precipitated precursor to convert to the desired magnetic soft phase (Fe$_3$O$_4$). The data presented, clearly show that there is a limited processing temperature window that produces the desired phase. Magnetic measurements reveal that the $(BH)_{max}$ of the SFO/$Fe_3O_4$ composite powders is 37% higher that the pure SFO hard phase, confirming that evasive goal of a rare earth free PM material can be realized using the procedure presented here.

## Acknowledgments

We thank Ms. S. Inoue, Mr. M. Manuel and Mr. R. Shah for help with sample synthesis and preparation. The support of this work from the Office of Naval Research with Dr. H. S. Coombe as program manager is most gratefully acknowledged.